\documentclass[preprint]{aastex}

\newcommand{\kms}{\mbox{${\rm\,km\,s}^{-1}$}}

\newcommand{\lum}{\mbox{${\rm\,erg\,s}^{-1}$}}
\newcommand{\flux}{\mbox{${\rm\,erg\,cm^{-2}\,s^{-1}}$}}

\newcommand{\apix}{\mbox{$\rm\,\AA\,pixel^{-1}$}}
\newcommand{\oii}{\mbox{$\rm\,[OII]\lambda3727$}}
\newcommand{\ha}{\mbox{$\rm\,H\alpha$}}
\newcommand{\hb}{\mbox{$\rm\,H\beta$}}
\newcommand{\mg}{\mbox{$\rm\,Mg_2$}}

\newcommand{\mgib}{\mbox{$\rm\,MgIb$}}
\newcommand{\ewoii}{\mbox{$\rm\,W_{[OII]}$}}
\newcommand{\ewha}{\mbox{$\rm\,W_{H\alpha}$}}
\newcommand{\ewhb}{\mbox{$\rm\,W_{H\beta}$}}

\newcommand{\img}{\mbox{$\rm\,I_{Mg_2}$}}
\newcommand{\ewmgib}{\mbox{$\rm\,W_{MgIb}$}}
\newcommand{\loii}{\mbox{$\rm\,L_{[OII]}$}}
\newcommand{\lha}{\mbox{$\rm\,L_{H\alpha}$}}
\def\gae{\mathrel{\hbox{\rlap{\hbox{\lower2pt\hbox{$\sim$}}}\hbox{\raise2pt\hbox{$>$}}}}}
\def\lae{\mathrel{\hbox{\rlap{\hbox{\lower2pt\hbox{$\sim$}}}\hbox{\raise2pt\hbox{$<$}}}}}

\begin{document}

\title{Evidence for Rapid Redshift Evolution of Strong Cluster Cooling Flows}

\author{R. Samuele} 
\affil{Northrop Grumman Aerospace Systems,
   One Space Park, Redondo Beach, CA 90278 }

\author{ B.R. McNamara} 
\affil{Department of Physics \& Astronomy, University of Waterloo  \& 
Perimeter Institute for Theoretical Physics, Waterloo, Ontario, Canada, \& Harvard-Smithsonian Center for Astrophysics, 60 Garden Street, Cambridge, MA 02138, USA}
\email{mcnamara@uwaterloo.ca}
\author{A. Vikhlinin}
\affil{Harvard-Smithsonian Center for Astrophysics, 60 Garden Street, Cambridge, MA 02138, USA}
\author{C.R. Mullis}
\affil{European Southern Observatory, Headquarters, Karl- \\ 
  Schwarzschild-Strasse 2, Garching bei Munchen D-85748, Germany}
\begin{abstract}

We present equivalent widths of the \oii\, and \ha\, nebular
emission lines for  77 brightest cluster galaxies (BCGs) selected from the 160 Square Degree $ROSAT$ X-ray
survey.   We find no \oii\, or \ha\, emission stronger than $-15$\,\AA\, or
$-5$\,\AA , respectively, in any BCG.  The corresponding emission line luminosities lie
below $\rm L \sim 6 \times10^{40}$~\lum\ , which is a factor of 30 below that of NGC1275
in the Perseus cluster.  A comparison to the detection frequency of nebular emission in
BCGs at $z \lae 0.35$ drawn from the  Brightest Cluster Survey  \citep{craw99} indicates that we should
have detected roughly one dozen emission-line galaxies, assuming the two surveys are selecting similar clusters 
in the X-ray luminosity range $10^{42}\rm ~erg~s^{-1}$ to $10^{45}~\rm erg~s^{-1}$.  The absence of luminous nebular emission (ie., Perseus-like systems)
in our sample is consistent with
an increase in the number density of {\it strong} cooling flow (cooling core) clusters between $\rm z=0.5$ and today. The decline in their
numbers at higher redshift could be due to cluster mergers and AGN heating.  

\end{abstract}


\clearpage

\section{INTRODUCTION}
Galaxy clusters are the largest virialized arrangements of dark matter
and baryons in the Universe.  Most of the baryons in clusters reside not 
in galaxies, but in hot, diffuse atmospheres that are easily detected in the X-ray band.
X-ray atmospheres are confined by deep gravitational wells,  so catalogs of X-ray-selected clusters are nearly insensitive to false concentrations of galaxies projected along the
line of sight.  Furthermore, X-ray luminosities provide a reliable estimate of cluster mass.  For these and other reasons,  X-ray cluster surveys are ideally suited for selecting and studying the systematic properties of clusters and their galaxy populations  (Rosati, Borgani, \& Norman 2002).

Brightest cluster galaxies (BCGs) reside near the X-ray centroid where a cooling flow (cool core) is often found (Fabian 1994).  A cooling flow is characterized by a bright, central cusp of X-ray emission from gas with a radiative cooling
time that is shorter than  $\rm 1\, Gyr$. BCGs centered in strong cooling flows usually harbor bright nebular emission,  star formation, and cold
molecular gas (Heckman et al. 1989; Edge 2001; Salom{\'e} \& Combes 2003; Donahue et al. 2000).   The connection between cold gas, star formation,  and cooling flows has been
strengthened by the recent discovery of a star formation threshold in BCGs. 
High star formation rates and bright nebular emission are found preferentially in BCGs centered in atmospheres with central cooling times shorter than $\sim 0.5$ Gyr (Rafferty et al. 2008, Cavagnolo et al. 2008, Hudson et al. 2010).   Therefore, the presence or absence of nebular emission in BCGs
signals the presence or absence of an active cluster cooling flow (see also Edwards et al. 2007
and Balogh et al. 2002).

We searched for nebular emission in a sample of 77 BCGs drawn from the 160 Square Degree  (160SD) $ROSAT$
X-ray cluster survey \citep{mul03, vik98}.  
We present nebular emission-line strengths of \oii\, and
\ha,  and we compare them to BCGs from the
Brightest Cluster Survey (BCS) \citep{eb98, craw99}.  Absorption lines
of \hb\, and \mgib\,, and the molecular band strength of \mg\, are presented.
No nebular emission or evidence of star formation is detected in any of
our BCGs,  which has interesting implications for the evolution of galaxy clusters
and cooling flows.

We assume a concordance $\rm \Lambda CDM$
cosmology with $\rm H_o=70\,km\,s^{-1}\,Mpc^{-1}$. 

\section{BRIGHTEST CLUSTER GALAXY SAMPLE AND OBSERVATIONS}

\subsection{The 160 Square Degree $ROSAT$ Cluster Survey}
The 160SD survey (Vikhlinin et al. 1998; Mullis et al. 2003)
yielded a catalog of  201 galaxy clusters detected by extended X-ray emission in 647
\emph{ROSAT PSPC} images.  The survey
covered roughly 160\,deg${^2}$ of the sky at high fluxes, 
approximately 80\,deg${^2}$ at a median flux of
${1.2{\times}10^{-13}}$\flux\, and 5 deg$^2$ at 3.7$\times10^{-14}$\flux.  The
survey includes targets at galactic latitudes greater than $30^\circ$ and with
hydrogen column densities below $6\times10^{20} \rm \,cm^{-2}$.  Regions
of 10$^\circ$ radius around the small and large Magellanic Clouds were excluded.  
The redshift range of the survey was $0.015\leq
z \leq1.261$, with a median redshift of $\rm \langle z \rangle = 0.25$.

\begin{figure}
\epsscale{0.5}
\plotone{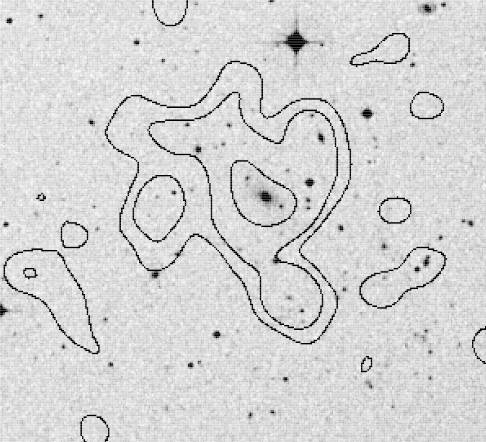}
\caption{An optical photometric image (10$^{\prime}\times10^{\prime}$)
of cluster RXJ0056.9--2740, with $ROSAT\, PSPC$ X-ray contours
overlaid.  The cluster is at a redshift of $\rm z = 0.116$ with an X-ray
luminosity of $\rm L_X = 1.6\times10^{44}$\lum.  The bright optical
source at the top of the image is a foreground star.  The brightest
galaxy located at the X-ray centroid of the cluster is clearly
discernible from the surrounding cluster members.}
\label{fig:bcg}
\end{figure}

Most extended X-ray sources away from the
Galactic plane are galaxy clusters \citep{mul03} that are distinguishable from nearby elliptical
galaxies and distant  quasars.
Sample selection is discussed throughly in  \citet{vik98}.
All cluster candidates were imaged in the optical band in order
to confirm their association with galaxy concentrations (McNamara et al. 2001).
BCGs were identified by overlaying $ROSAT$ X-ray contours onto the optical images (eg.,  Figure \ref{fig:bcg}), and 
spectra were taken for one or more
bright galaxies nearest to the X-ray centroid of each cluster  to measure redshifts (Mullis et al. 2003).  
When no obvious  BCG could be identified,  the galaxy with the brightest spectral continuum
was considered the BCG.  Of the 155 long slit spectra available from
the 160SD survey (45 of the 200 cluster redshifts were found in the
literature), 77 BCG spectra were sufficiently
sensitive for this study.  Most of the clusters with redshifts taken from the literature
are nearby, low luminosity systems whose redshifts were determined using absorption line spectra.   These systems
have been excluded from our study.

\begin{figure}
\epsscale{0.9}
\plotone{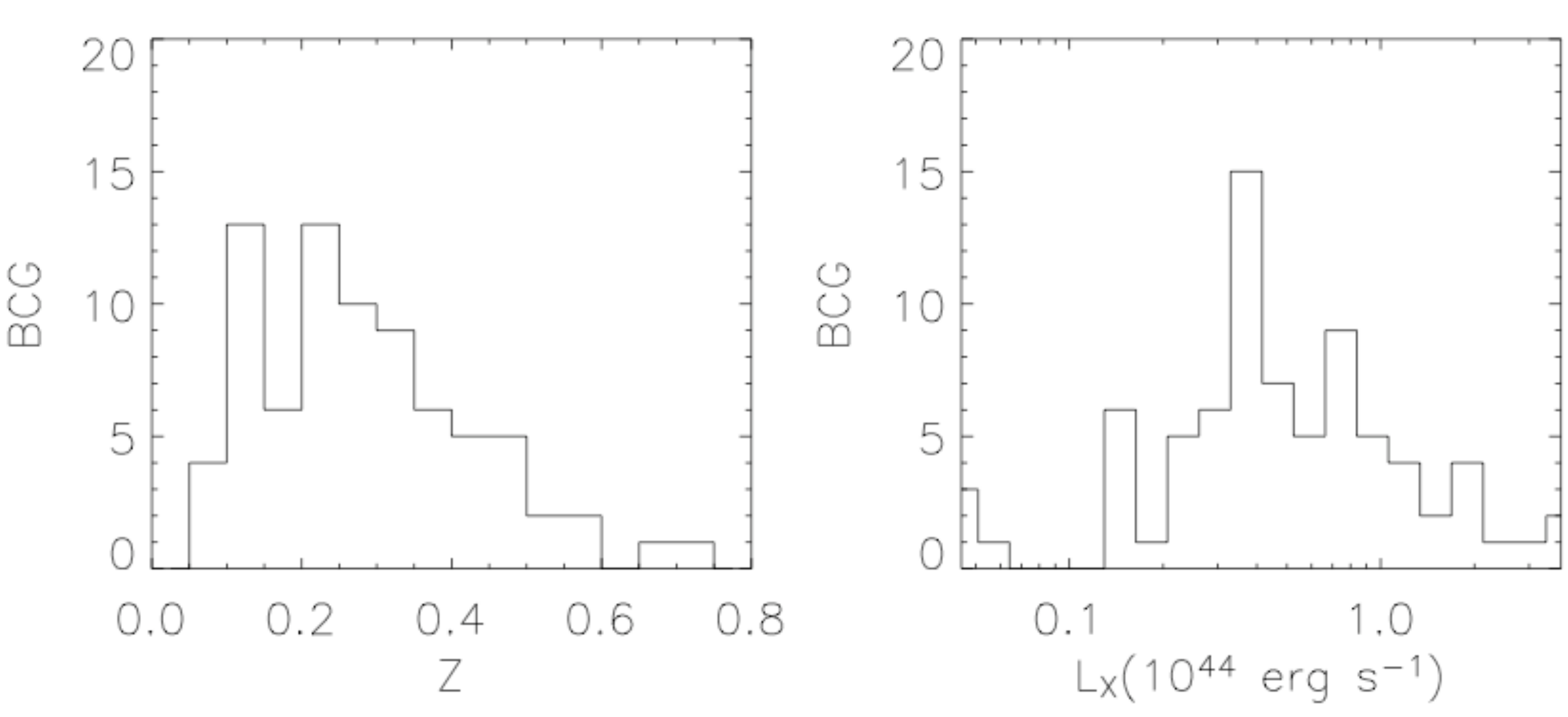}
\caption{(Left) The redshift distribution of the 77 BCGs represented
in this sample.  The sample ranges in redshift from 0.066---0.7, with
a median redshift of $\langle z \rangle = 0.256$ and a mean redshift of
$\rm \bar{z} = 0.286$.  (Right) The X-ray luminosity distribution on a
logarithmic scale of the host clusters for the sample.  The X-ray
luminosity for the clusters range from 0.04---4.27$\times10^{44}$\lum,
with a mean X-ray luminosity of $\rm \overline{L}_X = 7.2\times10^{43}$\lum\, and
a median of $\rm \langle L_X \rangle = 4.7\times10^{43}$\lum.}
\label{fig:sample}
\end{figure}

The redshift range of this sub-sample is $\rm z=0.066-0.7$. The redshift 
distribution is shown on the left panel of Figure
\ref{fig:sample}.   The mean and median redshifts are $\rm \overline{z} = 0.286$
and $\rm \langle z \rangle = 0.256$, respectively.  The X-ray
luminosity distribution of our clusters lies between
$\rm L_X=0.04-4.27\times10^{44}$\lum\ , as shown on the right panel of Figure
\ref{fig:sample}.  The sample is
weighted toward low luminosity clusters, with mean and median values of 
$\rm{\overline{L}_{X}=7.2\times10^{43}}$\lum\, and
$\langle\rm{L}_{X}\rangle=4.7\times10^{43}$\lum, respectively.

\subsection{Observations}
\begin{figure}
\epsscale{}
\plotone{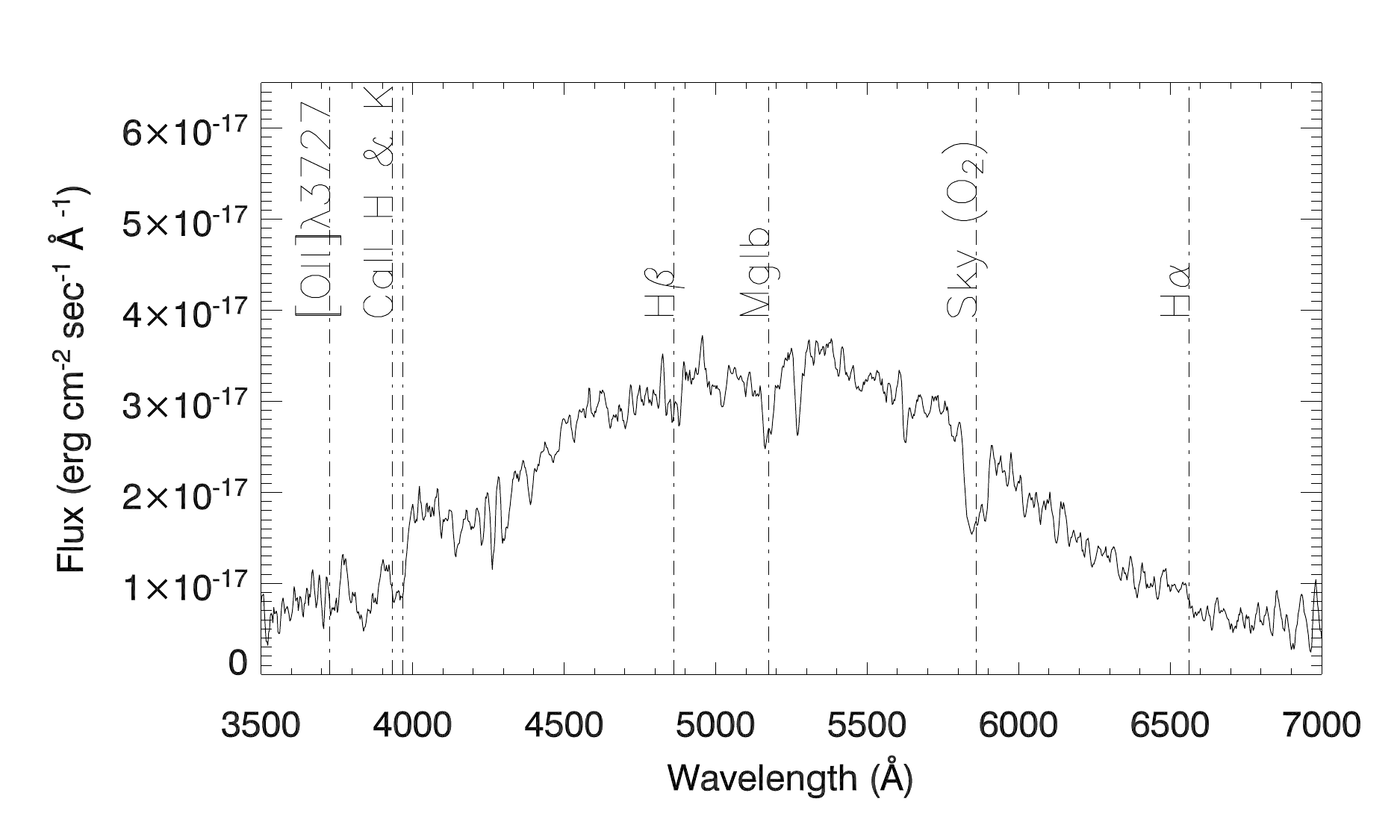}
\caption{A longslit spectrum of the BCG from cluster RXJ1343.4+5547 at
  z = 0.069 with $\rm L_X=4.0\times10^{42}\lum$, taken with the
  University of Hawaii 2.2\,m telescope.  The vertical dashed lines
  mark the location of emission and absorption features, with one
  large atmospheric O$_2$ absorption feature.  The spectrum has been
  smoothed using a boxcar average over 5 pixels}
\label{fig:spec}
\end{figure}

Three telescopes were used to collect spectra for this study: the
University of Hawaii (UH) 2.2\,m, Keck-II 10\,m and the Multiple
Mirror Telescope (MMT).  All 160SD survey spectroscopic observations
include the BCG and often two or three additional galaxies.
Spectra were gathered for 66 BCGs with the UH 2.2\,m telescope using a
1.8${^{\prime\prime}}$ slit width, an effective wavelength
coverage of 3800--9000\,\AA, and a 
a dispersion of roughly 3.6\,\AA\,pixel${^{-1}}$.  The spectra for six of
the most distant, $\rm z \geq 0.5$ BCGs were obtained with the Keck-II telescope.  Keck provided an
effective wavelength coverage of 5000--9000\,\AA~and a dispersion of
roughly 2.45\,\AA\,pixel${^{-1}}$ using a 1.5${^{\prime\prime}}$ slit width. Six BCG spectra were obtained
with the MMT using a slit width of 1.5${^{\prime\prime}}$, a wavelength coverage between
 3600--8400\,\AA, and a dispersion of roughly
1.96\,\AA\,pixel${^{-1}}$.  Figure
\ref{fig:spec} shows an example spectrum of a BCG observed with the UH
telescope that has been
smoothed with a boxcar average over 5 pixels.  The locations of
spectral features measured in this study are indicated.

\section{DATA REDUCTION AND ANALYSIS}

The data reduction was performed using IRAF.  The
spectra were de-biased, flat-fielded,  backgrounds were subtracted, and 
the wavelength scale was measured.  Details are given in  \citet{vik98} and
\citet{mul03}.  The spectroscopic redshifts were determined using the
4000\,\AA\, break and absorption features of Ca II H (3933.68\,\AA) and
K (3968.49\,\AA), the G band ($\sim$4300\,\AA), Mg Ib
($\sim$5175\,\AA), and Na Id (5889.97\,\AA), when detected.

\subsection{Line Equivalent Widths}
Line equivalent widths (W) were measured for \oii\, and \ha\,  in emission
and \hb\, and \mgib\, in absorption using the formula
\begin{equation}
 \rm  W = \int \, \frac{F_{c} \,- \, F_{\lambda}}{F_{c}} \, d\lambda,
\end{equation}

\noindent
where F$_c$ is the flux of the continuum and F$_{\lambda}$ is the flux
of the spectral line.  
We follow a sign convention where emission lines have negative equivalent widths, 
and absorption lines have positive equivalent widths.
The continuum was measured using adjacent sidebands redward and
blueward of the feature bandwidth.  A linear fit was applied across
the continuum sidebands. Feature bandwidths and continuum
sidebands are listed in Table\,\ref{table:widths}
\citep{ocon73,mc89,fab92}.  The sidebands were placed in close
proximity to the feature to avoid differential extinction and
continuum flux calibration uncertainty, and to avoid strong absorption
or emission features.  The full spectra have not been corrected for
line-of-sight extinction.  Local extinction corrections to
the emission line upper limit measurements have been applied.

Emission lines emerging from warm ionized gas are expected to
be Doppler broadened due to their motions in the center of the galaxy.  
Doppler broadened
\oii\, and \ha\, emission lines in cooling flow BCGs have velocity widths
lying between $\sim 200-700$\kms  \citep{heck89}.  Sampling dictates
that a minimum of 3 pixels be
included at the line position for each equivalent width measurement.
For our minimum dispersion of 3.42\apix, the minimum achievable
line sampling corresponds to  497\kms\, at \ha\, and 818\kms\, at \oii.  At our maximum dispersion
1.44\apix,  velocity widths approaching 264\kms\
for \ha\, and 450\kms\, for \oii\ can be resolved.  Our velocity width
coverage for \oii\, lies between $450-818$\kms\, and  $264-497$\kms\ for \ha\,.
Our spectra are  therefore reasonably well matched to the expected gas
velocity spread in these systems.

The estimated uncertainty in the equivalent width measurements
assume that the standard deviation of the best linear fit to the continuum
is equal to the uncertainty at the line center.
The uncertainties given in Table 2
are 1$\sigma$ standard deviation of the best linear fit.

\subsection{Spectral Indices}
The spectral index, $I$, is the ratio of the flux of the molecular band to
the interpolated continuum flux in magnitudes \citep{ocon73}.  We include a
measurement of $I$ for the \mg\, molecular band (MgH + Mg $b$) to
compare our BCGs to local giant elliptical galaxies \citep{dav93}.  The molecular
and continuum bandwidths for \mg\, are listed in
Table\,\ref{table:widths}.  To estimate the continuum, the average
flux of each continuum sideband was used to interpolate across the
molecular bandwidth \citep{ocon73, fab85}.  The spectral index was
measured using the following expression:

\begin{equation}
  I =
  -2.5\,log\left(\frac{F_{\lambda}(\lambda_2)}{F_{\lambda}(\lambda_1)+(F_\lambda(\lambda_3)-F_\lambda(\lambda_1))\left(\frac{\lambda_2-\lambda_1}{\lambda_3-\lambda_1}\right)}\right).
\end{equation}

In this expression,  $F_\lambda(\lambda_2)$ is the mean flux of the molecular
bandwidth, $F_\lambda(\lambda_1)$ and $F_\lambda(\lambda_3)$ are the
estimated mean fluxes of the continuum bandwidths, $\lambda_2$ is
the wavelength at the bandwidth center,  $\lambda_1$ and $\lambda_3$
are the wavelengths at the center of the continuum bandwidths
\citep{ocon73}.  The uncertainty in the spectral index measurement  was estimated by applying
the method of partial derivative error propagation and is presented  in Table 3 as the $1\sigma$ error estimate.

\section{RESULTS}
\subsection{Spectral Absorption Line Indices}

\begin{figure}
\epsscale{}
\plotone{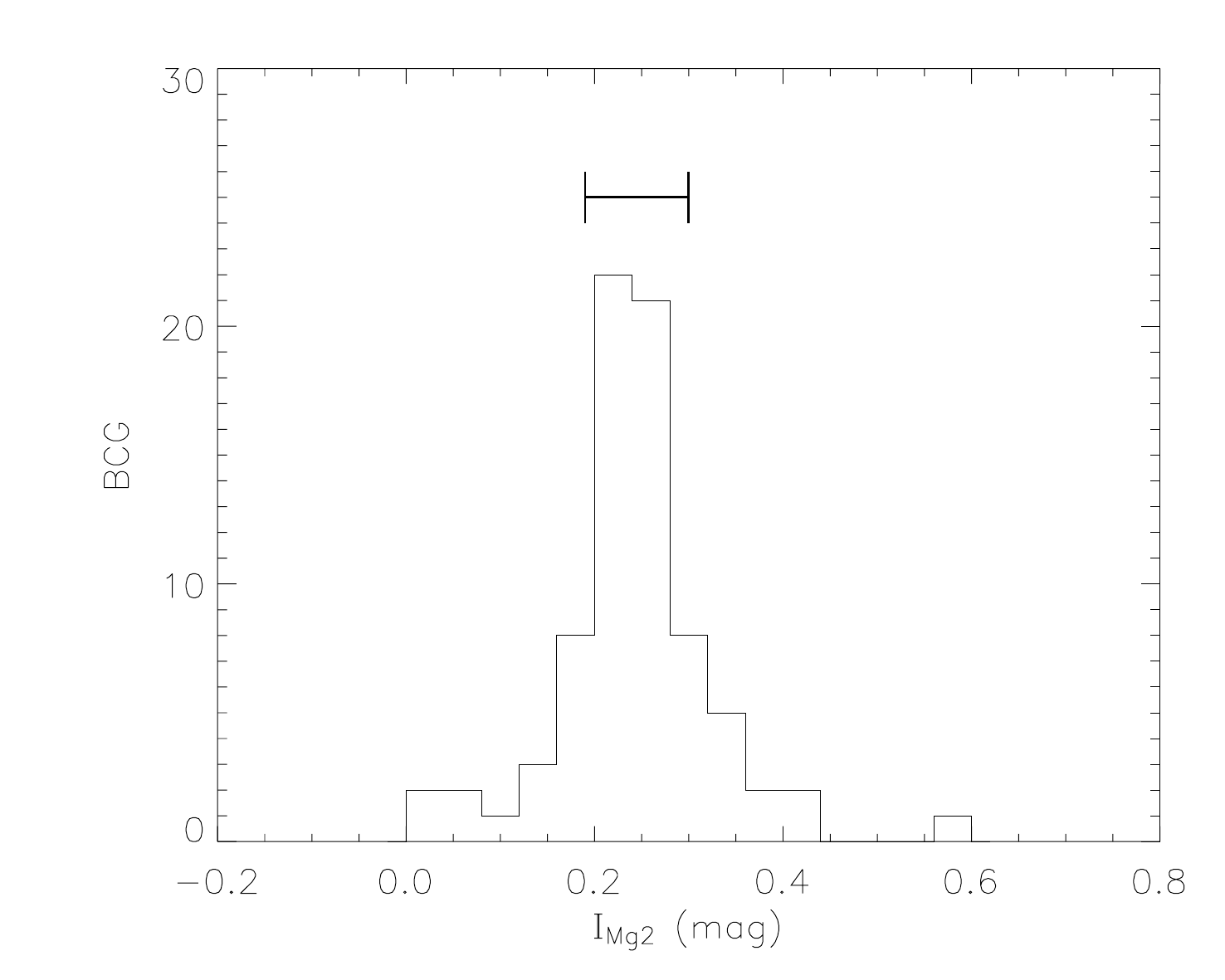}
\caption{A histogram of spectral indices for the Mg$_2$ molecular
 band for the sample of BCGs.  The boldface bar plotted above the
 histogram denotes the range of \mg\, values indicative of gE galaxies
 \citep{dav93}.  The mean of the sample ($\rm \overline{I}_{Mg2} =
 0.24 \pm 0.07\,mag$) falls central to the the range of local gEs.}

\label{fig:mg2}
\end{figure}

Though they were selected using similar X-ray criteria, the BCGs in this sample lie, for the most part, at higher redshifts than those in the Crawford et al. (1999) sample of BCGs drawn
from the $ROSAT$ Brightest Cluster Survey.  It is 
therefore worthwhile to compare key line indices for our galaxies to those for Crawford's galaxies and to nearby giant ellipticals \citep{dav93} with similar luminosities and
linear aperture sizes to identify any gross differences in their stellar populations.
In Figure\,\ref{fig:mg2} we present a histogram of the
distribution of \img\ indices for our sample.   Our sample yields a mean of $\rm
\overline{I}_{Mg2}=0.24\pm0.07\,mag$. In comparison, we found the mean Mg2 strength of the Crawford et al. (1999) sample
to be $0.28\pm 0.1$, which is consistent with the mean for our sample.  The horizontal bar in Figure 4 represents the
range of \img\ from the \citet{dav93} sample of local giant elliptical galaxies (gEs).  Of the BCGs plotted in Figure \ref{fig:mg2}, 65\% lie within the
range  of local gEs, and 97\% lie within 1$\sigma$\ of the mean.  Our Mg2 strengths are
also consistent with those for gEs in the Sloan Digital Sky Survey (Bernardi et al. 2003).

\begin{figure}
\epsscale{0.95}
\plotone{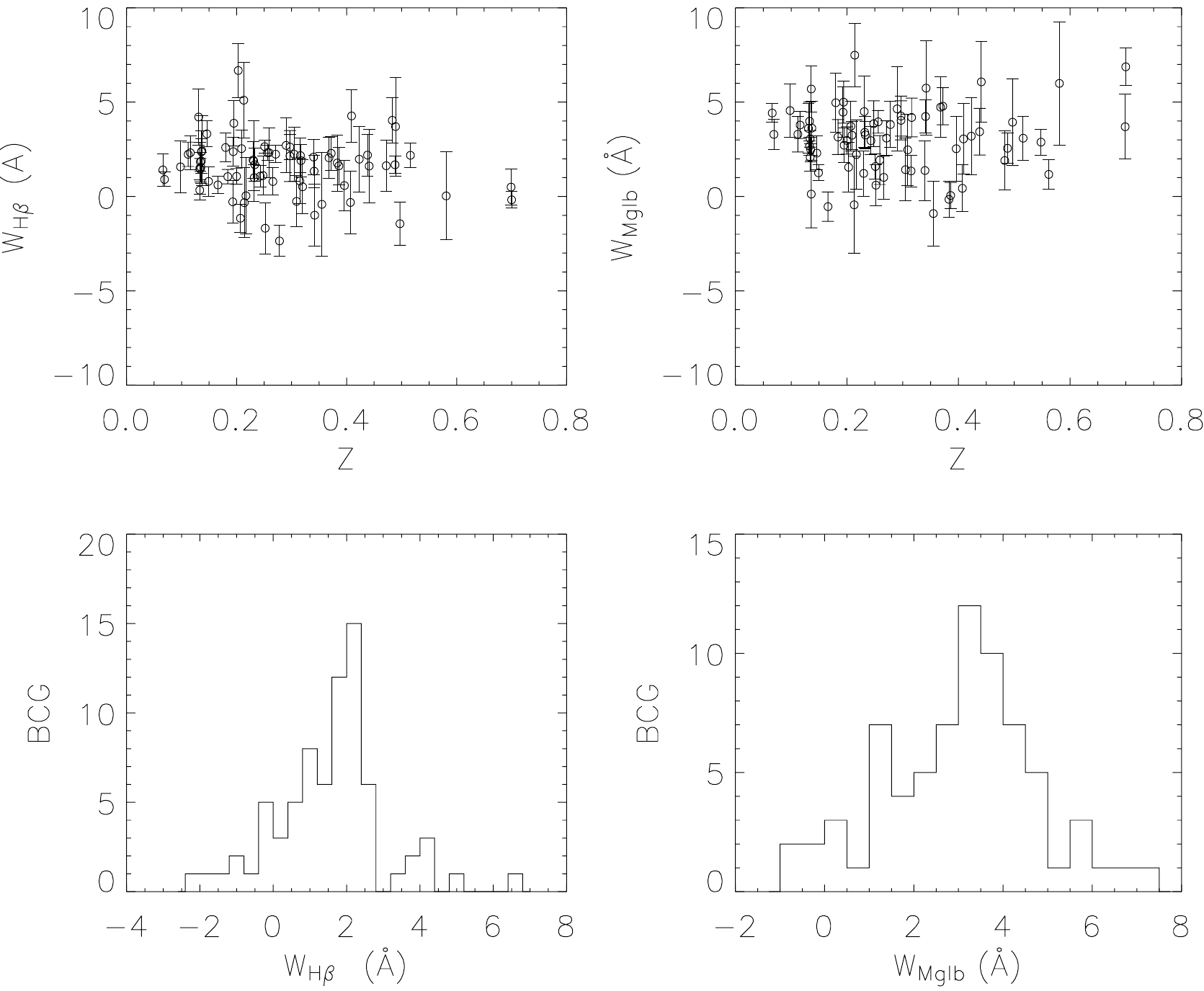}

\caption{(Upper-left) The equivalent width of \hb\, versus redshift
for the sample of BCGs.  Compared with local gEs, the BCGs show
similar \hb\, equivalent widths.  (Lower-left) The distribution of
\hb\, equivalent widths for the BCGs, centered about a mean
equivalent width of $\rm \overline{W}_{H\beta} = 1.57 \pm 1.52$\,\AA.
(Upper-right) The equivalent width of \mgib\, versus redshift, which
shows no significant gradient with redshift.  (Lower-right) The
distribution of measured \mgib\, equivalent widths versus redshift.
The adopted convention is that negative values indicate emission,
while positive values indicate absorption.}

\label{fig:abs}
\end{figure}

The lower-left panel of Figure\,\ref{fig:abs} shows the distribution
of \ewhb\, for our sample.  The range of \ewhb\, for the
\citet{dav93} sample is $\sim$0.4---2.8\,\AA, while the mean of our
sample is $\rm \overline{W}_{H\beta}=1.57\pm1.52$\,\AA. The BGCs fall
nicely within this range.  Considering the entire sample of BCGs for which
reliable measurements of \ewhb\ are available, $\sim$70\%
fall within the \citet{dav93} measured range, and $\sim$88\%
fall within 1$\sigma$\ of the distribution.  

An issue associated with this analysis concerns the effect of
metallicity gradients on our line strength measurements.  
It is well known that the mean stellar metallicity in gE galaxies and BCGs generally declines
with radius.  This implies that for a fixed slit
aperture, the observed spectrum of a distant galaxy extends to larger radii compared to
a nearby galaxy.  As a consequence, the stellar populations in the most distant galaxies
will appear to have lower
mean metallicities. The upper-right panel of Figure
\ref{fig:abs}\, shows the relationship between \ewmgib\ and redshift 
for the sample.  The plot of
\ewmgib\ versus $z$ shows no trend toward declining line strength in more
distant galaxies.  Therefore, we find
no discernible aperture-based metallicity bias within our sample. 

The upshot of this analysis is that we detect no gross anomalies in the
two absorption line indices studied here that would indicate
the presence of unusual stellar populations in our BCGs with respect
to our comparison sample and normal local giant elliptical galaxies.

\subsection{Lack of Nebular Emission}

Emission lines of \oii\, and \ha\, trace cooler gas that has been ionized by hot, young
stars and an active nucleus.  The  \oii\, emission-line is sensitive to
low levels of ionized gas because of its
intrinsic strength and high contrast against the declining stellar UV continuum.  However, measuring \ewoii\, can be
challenging given the local sensitivity to metallicity and the declining sensitivity of CCD
detectors in the near UV \citep{bal02}.  

Because the signal-to-noise ratios of
our spectra were optimized for redshift measurements and not for the detection 
of {\it weak} line emission, we shall restrict our discussions to galaxies
of moderate to high emission line strengths, which would be easily detected.    
We have chosen to present our upper limits primarily in the form of equivalent widths, rather than 
line luminosity for practical reasons.  First, equivalent width is a differential quantity that can be measured
accurately and independently of the flux calibration.  Because our spectra were taken over several years and
with three different instruments, equivalent width measurements avoid any systematic differences in flux calibration
between the various observing runs.  Second, Crawford et al. (1999) presents line luminosities for the objects they
detected, but they do not present upper limits. Therefore, a direct comparison between the two
samples in luminosity would not be feasible.  Furthermore, nebular line emission is often spatially extended, and comparisons
between line luminosities would be hampered by unknown aperture corrections.  Equivalent widths, being differential
quantities are less affected.  Crawford et al. (1995) presented equivalent widths for a large number of their objects, which allowed us to convert their line luminosities
to equivalent widths for the purpose of this comparison.  

\begin{figure}
\epsscale{0.95}
\plotone{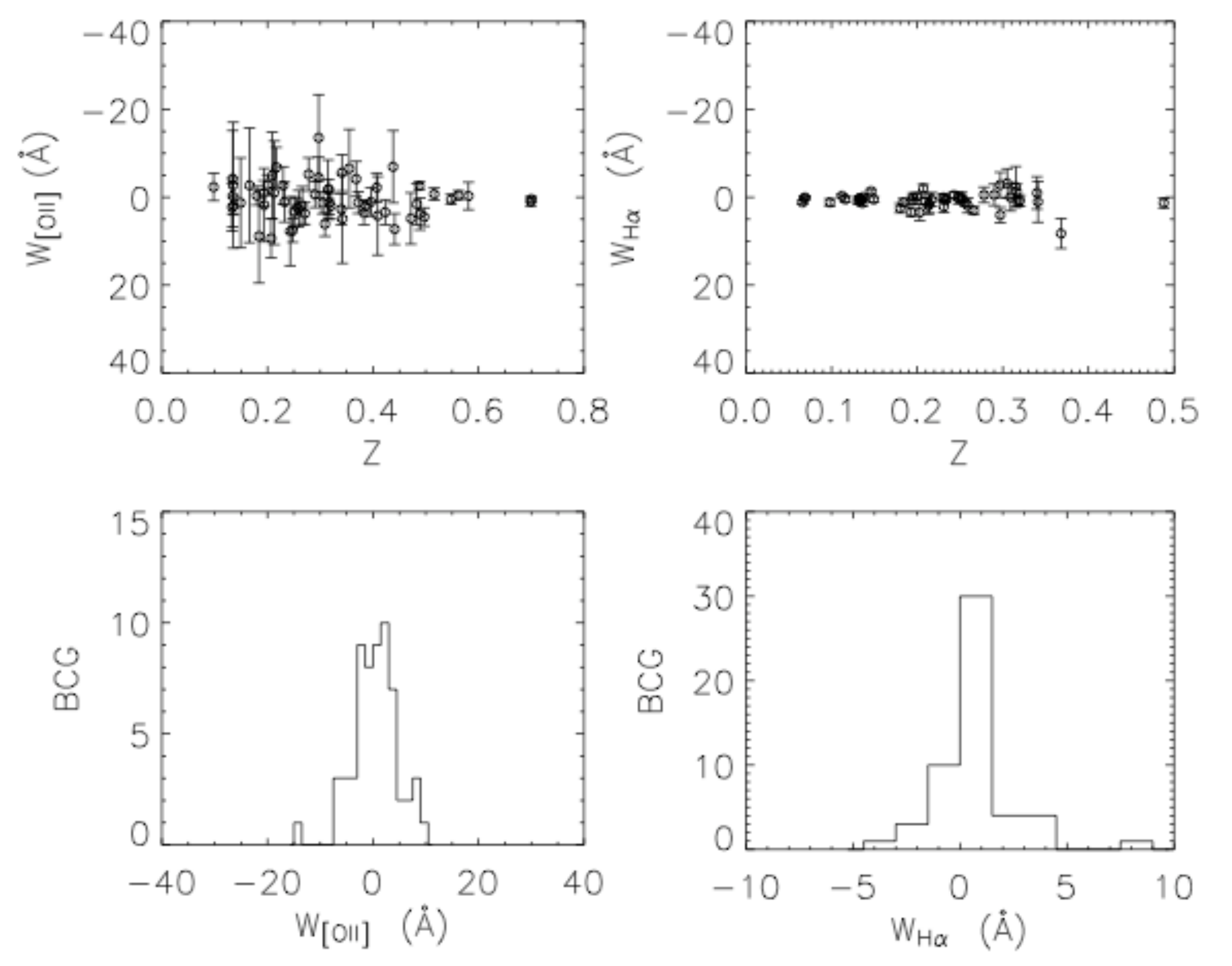}

\caption{(Upper-left) The equivalent width of \oii\, versus redshift
for the BCGs of this sample.  The mean of \oii\, equivalent
width is $\rm \overline{W}_{[OII]}=0.56 \pm 4.39$, and no significant
\oii\, emission stronger than $\sim-15$\,\AA\, was detected in the
sample. (Lower-left) The distribution of \oii\, equivalent widths
among the BCGs.  (Upper-right) The equivalent widths of \ha\, versus
redshift, where the mean is $\rm
\overline{W}_{H\alpha} = 0.64 \pm 1.76$\,\AA, and no significant \ha\,
emission stronger than $\sim-5$\AA\, has been detected in the
sample.  (Lower-right) The distribution of \ha\, equivalent widths
among the BCGs.  The adopted convention is that negative values
indicate emission, while positive values indicate absorption.}

\label{fig:em}
\end{figure}

Figure \ref{fig:em}\, shows the equivalent widths of  \oii\  and \ha\ as
a function of redshift.   For objects below redshift $0.2$, the \oii\,
feature falls in a noisy region near the edge of the
wavelength coverage of the spectrum.  For this reason, 16 \ewoii\,
measurements were excluded from Figure \ref{fig:em}.  Similarly,
measurements of \ewha\, for nine galaxies with redshifts above $ 0.35$ have
been omitted from Figure \ref{fig:em} because they fall outside of
the wavelength coverage of the spectra. An additional 13 measurements were
excluded because of high noise or their coincidence with a
sky line.  The \oii\, and \ha\, features are complementary
tracers of nebular emission.  One or both were
measured for each BCG.

The upper-left panel of Figure \ref{fig:em}\, shows \ewoii\, as a
function of redshift.   The mean equivalent width is $\rm \overline{W}_{[OII]}=0.56\pm 4.39$. No
emission line stronger than  $3\sigma$ of the mean, or  $\sim-15$\,\AA\ is detected.
The largest  \oii\, emission line strength is $\rm W_{[OII],MAX}=-13.6 \pm 9.8$\AA  ,
but this feature is probably spurious.
The upper-right panel of Figure \ref{fig:em}\, shows \ewha\, as a
function of redshift.  The increasing uncertainty in \ewha\, with redshift
results from increased noise in the \ha\,
spectral region for objects at higher redshifts.   The mean for the sample is $\rm
\overline{W}_{H\alpha}=0.64\pm1.76\,\AA$.  Again,  no significant \ha\
emission above  $3\sigma$ of the mean, or $\sim-5$\,\AA\, is detected.
In fact, most show \ha\, in absorption.

One sigma upper limits to the line luminosities were calculated for each of the
\oii\, and \ha\, emission lines.  Six uncalibrated MMT
spectra were excluded.  Emission-line luminosities
were corrected for line-of-sight extinction
following \citet{card89}.
The average 1$\sigma$ line upper limits  for \oii\, and \ha\, are
$\rm \overline{L}_{[OII]}=6.1\times10^{40}$\lum\, and $\rm
\overline{L}_{\rm H\alpha}=3.8\times10^{40}$\lum, respectively.  The median
values are $\rm \langle L_{[OII]} \rangle = 3.7\times10^{40}\lum$ and
$\rm \langle L_{H\alpha} \rangle = 2.1\times10^{40}\lum$.  Table
\ref{tab:tabem} lists the \loii, \lha, and corresponding equivalent
widths, the cluster redshifts, and their X-ray luminosities.
Table \ref{tab:tabab} lists the equivalent widths 
for \hb\, and \mgib, as well as the \mg\, spectral indices for each
cluster.  No significant \oii\, emission stronger than $\sim-13$\,\AA\, or \ha\,
emission stronger than $\sim-5$\,\AA\, has been detected in any
BCG in our sample. 

The absence of strong line emission in any 160SD BCG is surprising.
The strongest emission lines seen in cooling flow BCGs 
have \oii\  equivalent widths  that are significantly greater than our upper limits.
For example, \oii\ reaches an equivalent width of
$-75.9$\,\AA\ in the Abell 1795 BCG and $-39.6$\,\AA\  in the Abell 2052 BCG \citep{mc89}.  The
\ewoii\, measurements of
\citet{craw95} yielded values of $-58.0$\,\AA\ for RXJ0439.0+0520 and  $-23.5$\,\AA\ for
the A291 BCG.  Objects of this nature are clearly missing from our sample.

\section{Comparison to The $ROSAT$ Brightest Cluster Survey}

The Brightest Cluster Survey (BCS) is a compilation of 201 clusters selected
from the $ROSAT$ All Sky Survey (RASS), the majority of which have 
X-ray luminosities exceeding $\rm L_X = 10^{44}~erg\,s^{-1}$.  Most lie below redshift $\rm
z=0.3$ \citep{eb98}.  \citet{craw99} searched for emission-lines in a sample of 177
BCGs in clusters selected from the BCS to an unabsorbed X-ray flux limit
of $\rm7.9 \times10^{-12}\,erg\,cm^{-2}\,s^{-1}$.  Their sample yielded
a detection fraction of 27\% in \ha\  emission, while 32\% show emission
in other lines such as [NII]$\lambda\lambda$6548,6584.  

\begin{figure}
\epsscale{0.93}
\plotone{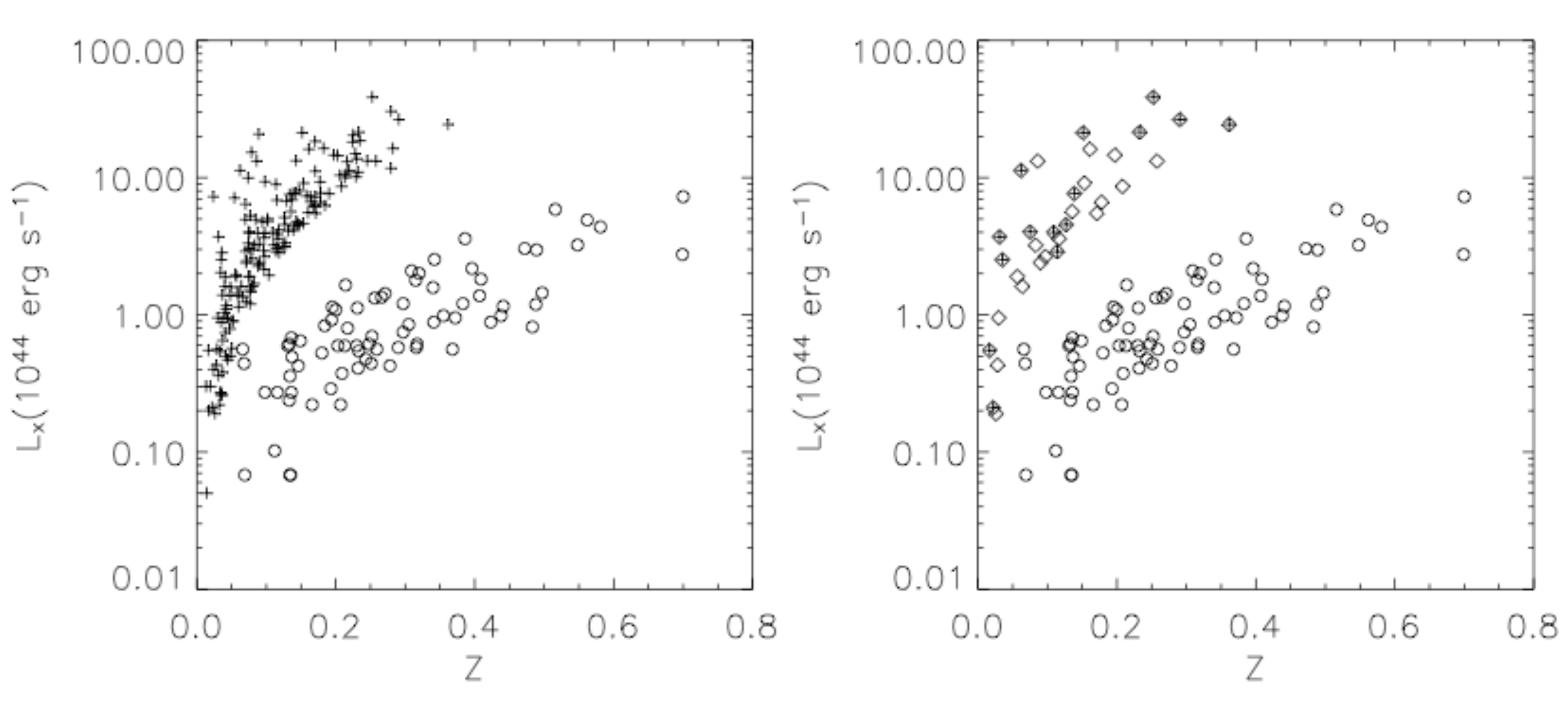}

\caption{(Left panel) A comparison of the BCS sample of \citet{craw99},
marked with '+', and this sample, marked with '$\circ$', with X-ray
luminosity plotted as a function of redshift for the BCGs.  (Right
panel) The emission-line BCGs of the BCS sample plotted with cluster
X-ray luminosity as a function of redshift.  The BCGs that yield \ha\,
emission stronger than $-10$\,\AA\, are marked with a '$\Diamond$', while
BCGs that yield \oii\, emission stronger than $-20$\,\AA\, are filled with
a '+'.  The BCGs of this sample are also plotted and marked with
'$\circ$'.}

\label{fig:fluxcomp}
\end{figure}

The left panel of Figure \ref{fig:fluxcomp}
compares the X-ray luminosities of Crawford's BCS clusters (indicated as
``+") to our  sample (indicated by  ``$\circ$") versus redshift.  We have multiplied the 160SD X-ray
luminosities in this figure by the factor 1.7 to reflect the passband difference between  BCS
luminosities and ours. 
Figure \ref{fig:fluxcomp} shows that the highest BCS X-ray luminosities at roughly $\rm
L_{X}=4\times10^{45}\lum$ exceed the most luminous
160SD cluster by about five times.   Nevertheless, there is substantial overlap in the X-ray luminosities
of clusters from both surverys.   Apart from a few low luminosity clusters, we find that
at a fixed X-ray luminosity, the 160SD clusters lie at higher redshifts
compared to their BCS counterparts.  

The right panel of Figure \ref{fig:fluxcomp} shows only the BCGs 
from Crawford's sample with line emission at levels above our
detection threshold.  Crawford et al. found 
a roughly constant detection frequency of line emitting BCGs with
redshift and luminosity \citep{craw99}.  Accordingly, this diagram shows many line
emitting BCGs in clusters with comparable X-ray luminosities to 160SD clusters.
Therefore, the absence of line emission in 160SD BCGs is not
entirely due to their association with lower luminosity clusters. 
One distinguishing characteristic is that BCS clusters of a given X-ray luminosity lie at lower 
redshifts compared to their 160SD counterparts.   We explore this below.

\begin{figure}
\epsscale{}
\plotone{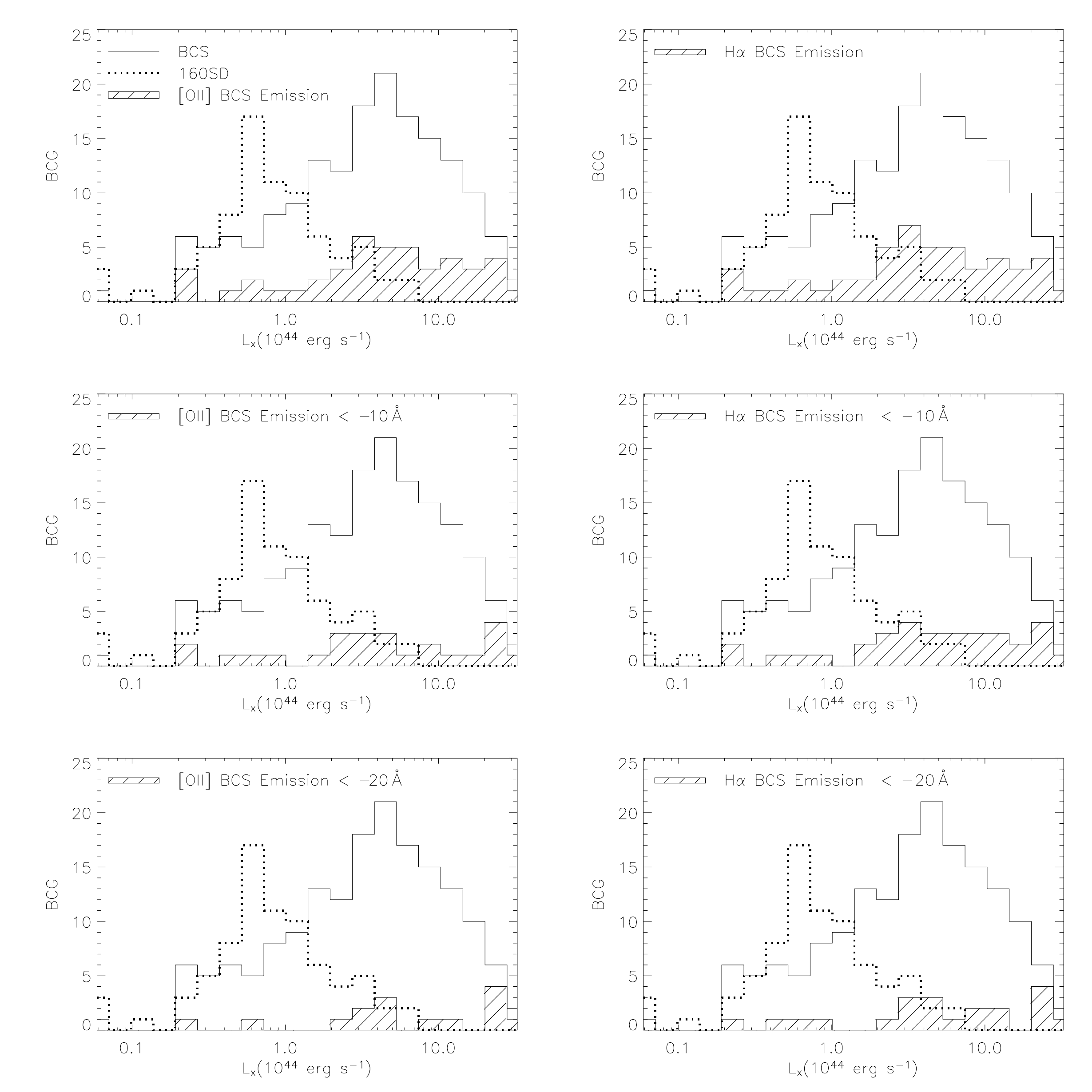}
\caption{A set of histograms that show the X-ray
luminosity distribution of BCGs for the Brightest Cluster Sample
(BCS), this sample (labeled as 160SD), and the BCS emission line
galaxies.  For comparison purposes, the X-ray luminosities of this
sample have been corrected to the RASS bandwidth [0.1--2.4 keV].  The
left and right columns are designated for BCGs that show \oii\, and
\ha\, emission respectively.  The second row shows detections stronger than
$\sim-10$\,\AA\, for \oii\, and \ha\,, while the third row shows detections
stronger than $\sim-20$\,\AA.}
\label{fig:comp}
\end{figure}

The histograms in Figure \ref{fig:comp} show the X-ray
luminosity distributions of the BCS and 160SD clusters discussed here.  
BCS clusters harboring emission line BCGs are hatched.  The 160SD
X-ray luminosities have been corrected to the RASS bandpass.
The  \oii\ and \ha\ detections from the BCS are shown in the left and right columns, respectively.
The top row summarizes all detections.  The second row shows detections stronger
than $\sim-10$\,\AA\, for \oii\, and \ha\, respectively, while the
bottom row shows detections stronger than $\sim-20$\,\AA.  

In order to make this comparison, we converted 
Crawford's emission line luminosities to equivalent widths.
 The conversion was made using several BCGs with line fluxes from
 \citet{craw99} found in common with \ewoii\, measurements from \citet{mc89}.
 The average ratio of equivalent width to flux was multiplied by the
  \oii\, line flux reported by \citet{craw99} for each line emitting
  BCG. These ratios are  5.2$\pm$2.6 for \oii\  and 9.7$\pm$4.5 for \ha .
Of the 177 BCS clusters presented by \citet{craw99}, 10 were
omitted from Figure \ref{fig:comp} because their  X-ray
luminosities were unavailable to us.  Of these,  four BCGs have \ha\,
in emission and two have \oii\, in emission.  Excluding these objects, 
the BCS contains 44 BCGs with \oii\, in emission and 49
with \ha\, in emission.  Fifty seven percent of the \oii\,
emission-line BCGs and 67\% of \ha\, emission-line BCGs have
equivalent widths stronger than $\sim-10$\,\AA.  Thirty four percent  of \oii\,
emission-line BCGs and 43\% of
\ha\, emission-line BCGs show equivalent widths stronger than
$\sim-20$\,\AA.

The nebular emission detection thresholds for the BCS shown in Figure \ref{fig:comp} 
lie comfortably above ours.  Such objects, if present in our sample,
would have been detected.   
Of the entire sample of 177 BCS BCGs  from \citet{craw99}, 121 fall within
the X-ray luminosity range of this sample.   Within this luminosity range,
7\% of  BCS BCG have \oii\, emission stronger than $\sim-20$\,\AA\, and 16\% have
\ha\, emission stronger than $\sim-10$\,\AA.  No 160SD BCG reveals nebular
emission at this level.

The right panel of Figure \ref{fig:fluxcomp}\, shows the X-ray
luminosities of the 160SD BCGs and BCS BCGs harboring nebular emission above our thresholds
versus redshift.  The BCGs revealing only \oii\,
emission are marked with a ``$\Diamond$"; those with both
\oii\, and \ha\, emission are indicated as ``+".  This figure shows
that at a given redshift, BCS clusters have higher average X-ray luminosities.

About 30\% of BCS BCGs with cluster X-ray luminosities
greater than $\rm L_X=10^{45}$\lum have emission lines.  
The fraction falls to 18\% within the X-ray luminosity range
$\rm 10^{44}\lum>L_X\geq10^{45}\lum$.
Similarly, 16\% of clusters with X-ray
luminosities within the range $\rm 10^{42}\lum>L_X\geq10^{44}\lum$ have
emission lines.   These figures show that while
higher luminosity clusters show a higher frequency of emission line BCGs, 
the lower average X-ray luminosity of the 160SD sample 
cannot alone explain the absence of strong line emission.  

Table \ref{tab:emcomp} summarizes the detection statistics for both surveys.
Assuming both samples are equally likely to have BCGs with bright line emission within the overlapping
luminosity range,  the table shows that we should have detected line emission in roughly one dozen 160SD BCGs
with H$\alpha$ equivalent widths greater than $-10$ \AA , and with [O II] widths exceeding $-20$ \AA . The probability
of having detected one or fewer emission line BCGs by chance is 
approximately one in ten thousand.  Eight percent of BCS clusters within our X-ray luminosity range host BCGs with line equivalent widths stronger than $-20$ \AA.
We would have expected to detect about 6 such systems in our sample (Figure 8, panel 3), yet we detected none.  The likelihood of
having detected one or fewer such system by chance is less than $2\%$.  Thus, the absence of line emitting BCGs in the 160SD survey
is statistically significant at a high level.  

\subsection{Potential Systematic Errors}

A potential systematic effect that could bias our equivalent widths downward is beam dilution owing to larger projected linear
aperture sizes with increasing redshift.  Crawford et al. used a slit width of  approximately 1.2  arcsec, which is somewhat smaller than ours.  Crawford's clusters
lie on average at lower redshifts,  so that their slit width projects over a several kpc on the galaxy.  In comparison,  our slits subtended a  linear 
width of between $6-9$ kpc between redshifts of 0.3 and 0.5.  Furthermore, Crawford integrated along the slit  out to a typical linear distance of about 10 kpc,
while we integrated along the slit to the approximate location where the
galaxy brightness falls below the sky brightness, which is $3-5$ times larger in radius. Because of the larger fraction of galaxy light in our
aperture compared to Crawford's, our equivalent may be diluted with respect to theirs.

We have estimated the magnitude of the effect by modeling the light profile of our galaxies with
a Hubble law, ie., stellar surface brightness declining radially as $r^{-2}$, and integrating the stellar light beyond the central 10 kpc down to the sky limit.   We found that the effect would 
bias our equivalent widths downward by at most $30\%-40\%$ with respect to Crawford's.  In other words,  it is possible that an object with
$H\alpha$ equivalent width of $-7$\AA~ would fall below our $-5$\AA~ cutoff because of aperture dilution.  However, the {\it detection} of stronger
features is not in jeopardy.  In fact, an upper limit on the magnitude of the effect can be evaluated directly by scanning the line luminosity upper 
limits for the most distant objects shown in Table 2,  
which would show the largest effects.  These upper limits lie below $10^{41}~\rm erg~s^{-1}$.
In comparison, the objects that are putatively missing from our sample would have line luminosities lying between $10^{41}~\rm erg~s^{-1}$ and
$10^{43}~\rm erg~s^{-1}$.  They would easily have been detected.  In fact, Crawford's  smaller aperture will have missed flux  from spatially extended emission line systems that we would have captured.  Many emission line systems in cooling flows, such
as Perseus,   extend  30 kpc or more from the nucleus (Heckman et al. 1989, O'Dea et al. 2010).   
Therefore, a dilution effect will be most pronounced in weak emission line galaxies 
with small, unresolved nebulae, which we are not concerned with in this paper.

While the detectability of strong emission lines is not at issue, our primary concern is whether the X-ray
selection functions in the 160SD and RASS surveys are oppositely biased toward the non detection and detection of cooling flow clusters
in each respective survey. In a study of 10 BCGs from the 160SD survey with X-ray
luminosities below $\rm L_X=4\times10^{43}\lum\,$ [0.1--2.4 keV],
\citet{bal02} also failed to detect line emission in any BCG.   They argued that the 160SD
survey is biased against clusters with strong cooling flows because
only extended sources at a faint X-ray flux limit were sought.  This interpretation is
incorrect.  If the 160SD survey is biased, we would expect it to contain an excessive
number of cooling flows.  The central  $\sim 100$ kpc or so of cooling flow clusters are
brighter than normal clusters.   They are well resolved in the 160SD survey,
and are thus easier to detect than non-cooling flow clusters (Vikhlinin et al. 1998, Burenin et al. 2007).  Simulations have shown that the detection probability declines precipitously
only for clusters harboring X-ray-bright AGN near their centers, and only in the rare event that the AGN outshines the cluster emission (Burenin et al. 2007).  Such clusters are rare at
these redshifts.

Figure 7 shows that clusters of a given X-ray
luminosity in the BCS sample lie at systematically larger redshifts compared to  their 160SD survey counterparts.  
If the lack of emission lines in the 160SD survey is a physical effect and is not
due to selection bias,  it appears to be related to redshift.  This interpretation implies
that at a given X-ray luminosity, there are fewer well established {\it strong} cooling flows at
higher redshift, suggesting rapid evolution between  redshift one-half and today.

Note that our relatively high detection threshold is insensitive to the existence of 
weak cooling flows in the 160SD survey.  Roughly  15\% of BCGs selected from the Sloan and
NOAO Fundamental Plane cluster surveys (Edwards et al. 2007) have detectible line emission.
This fraction rises to near unity for bona-fide cooling flow clusters.  However, the
H$\alpha$ equivalent widths of most are weaker than our $-5$\AA\  threshold.
Nevertheless, 28 of their 328 BCGs (8\%) drawn from the Sloan survey have H$\alpha$ equivalent
widths stronger than $-5$\AA, and  all of them
lie at redshifts at or below $\sim 0.1$.  A direct comparison between our survey
and the Edwards et al. (2007) survey is infeasible because of the different selection
methods.    However, it is clear that were similar BCGs present in our survey, 
we would have detected a substantial fraction of them. The
absence of any BCG with strong line emission in our sample bolsters
the case for an evolutionary effect.    

\section{DISCUSSION}

We searched for nebular emission in a sample of 77 BCGs from the 160SD
X-ray cluster survey.   No \oii\, or \ha\, emission stronger than  $-15$ \AA~ and $-5$\AA, respectively, 
was detected in any BCG.
A comparison between our sample and the BCS clusters of similar X-ray 
luminosity suggests that we should have
detected roughly one dozen emission-line BCGs stronger than our detection threshold.
160SD clusters of a given X-ray luminosity lie at systematically higher redshifts
compared to BCS clusters of the same luminosity.  This may indicate an  increase
in the number density of {\it strong} cooling flows between $z=0.5$ and today.  Because of
our relatively high detection thresholds, our result does not imply that cooling flows
don't exist at the redshifts surveyed in the 160SD survey (see Bauer et al. 2005).  However,
we find a significant decrease in the number of {\it strong} cooling flows like the Perseus
Cluster.  Similar conclusions were reached using independent methods 
by Santos et al. (2008), Vikhlinin et al. (2007) and by  Voevodkin et al. (2010), who also found fewer  cooling flow clusters at higher redshifts.  

The decline may be related to the destruction of cooling flows by mergers, which occurred more
frequently than they do today (Santos et al. 2008).  Weaker mergers may also disturb
X-ray atmospheres and miscenter the BCG, which could prevent a
strong cooling flow from forming.   AGN feedback is thought to be responsible for quenching cooling flows in nearby clusters (Peterson \& Fabian 2006; McNamara \& Nulsen 2007), but this is
less clear in higher redshift clusters lacking the signatures of cooling flows.  However,  in a study
of radio AGN located in the 400 Square Degree X-ray cluster survey (Burenin et al. 2007),
Ma et al. (2011, in preparation)  found that the power injected by radio sources in galaxies
lying within 250 kpc of cluster centers  is significant compared to the power radiated by
the hot atmospheres.  The selection function of the 400 Square Degree Survey is
nearly identical to the 160 SD survey, but it covers a much larger area
of the sky.  The Ma et al. study suggests that AGN are able to deposit enough energy
in these clusters to delay or prevent a massive cooling flow from forming at redshifts of
a few tenths or so. 

McNamara and Samuele would like to acknowledge Ohio University's Astrophysical Institute for its hospitality while much of the data analysis was performed in pursuit of Samuele's masters
thesis.  McNamara acknowledges conversations with Alastair Edge, Megan Donahue,
Paul Nulsen, Andy Fabian, and a generous grant from Canada's Natural Sciences and Engineering 
Research Council.

 

\clearpage

\include{FIG_abs}

\include{FIG_comp}

\begin{deluxetable}{cccc}
\tablewidth{0pt}
\tablecolumns{4}
\tablecaption{The feature bandwidths and continuum sidebands used to measure line strengths.}
\tablehead{\colhead{Feature} & \colhead{Line Center (\AA)} &\colhead{Bandwidth (\AA)} & \colhead{Continuum Sidebands (\AA)}}
\startdata
  [OII]             &3727                 &3721.00--3732.00   &3692.00--3722.00, 3732.00--3762.00\\
  H${\alpha}$       &6562                 &6557.00--6568.00   &6527.00--6557.00, 6567.00--6597.00\\
  H${\beta}$        &4861                 &4849.50--4877.00   &4829.50--4848.25, 4878.25--4892.00\\
  MgIb              &5175                 &5162.00--5193.25   &5144.50--5162.00, 5193.25--5207.00\\
  Mg${_{2}}$\tablenotemark{\dagger} & --                  &5156.00--5197.25   &4897.00--4958.25, 5303.00--5366.75\\
\enddata
\label{table:widths}
\tablenotetext{\dagger}{Mg$_2$ is a molecular bandwidth measured using a spectral index in magnitudes.}
\end{deluxetable}

\begin{deluxetable}{ccccccc}
\rotate
\tablecolumns{5}
\tablewidth{0pt}
\small


\tablecaption{The emission line strengths and upper luminosity limits for the BCGs}
\tablehead{\colhead{Object} & \colhead{L$_{\rm X}$} & \colhead{z} &
\colhead{W$_{\rm [OII]}$} & \colhead{L$_{\rm [OII]}$\tablenotemark{\Lambda}} &
\colhead{W$_{\rm H\alpha}$} & \colhead{L$_{\rm H\alpha}$\tablenotemark{\Lambda}} \\ \colhead{} &
\colhead{($10^{44}\,\rm erg\,s^{-1}$)} & \colhead{} & \colhead{(\AA)} &
\colhead{($<10^{40}\,\rm erg\,s^{-1}$)} & \colhead{(\AA)} &
\colhead{($<10^{40}\,\rm erg\,s^{-1}$})}
\startdata

RXJ0041.1--2339\tablenotemark{*}& 0.06& 0.112& ...& ...&  -0.34 $\pm$\,
  0.31&   0.85 \\ 
RXJ0050.9--0929& 0.64& 0.200&  -2.82 $\pm$\,  2.28&   3.97&  -0.11 $\pm$\,
  0.48&   2.40 \\ 
RXJ0056.9--2213\tablenotemark{*}& 0.16& 0.116& ...& ...&   0.47 $\pm$\,
  0.66& ... \\ 
RXJ0110.3+1938 & 0.36& 0.317&   1.25 $\pm$\,  2.72& ...&   0.83 $\pm$\,
  1.29& ... \\ 
RXJ0122.5--2832& 0.78& 0.256&   2.24 $\pm$\,  1.91& ...&   1.27 $\pm$\,
  0.73& ... \\ 
RXJ0124.5+0400 & 0.34& 0.316&  -1.85 $\pm$\,  2.26&   3.59&  -1.36 $\pm$\,
  1.65&   8.36 \\ 
RXJ0142.8+2025\tablenotemark{\dagger} & 0.84& 0.271&   3.68 $\pm$\,  2.47& ...& ...& ... \\ 
RXJ0144.4+0212\tablenotemark{sky} & 0.13& 0.166&  -2.70 $\pm$\, 13.02&   0.77& ...& ... \\ 
RXJ0159.3+0030\tablenotemark{\dagger} & 2.11& 0.386&   1.95 $\pm$\,  1.23& ...& ...& ... \\ 
RXJ0206.3+1511 & 0.36& 0.248&   7.56 $\pm$\,  2.66& ...&   0.17 $\pm$\,
  1.29& ... \\ 
RXJ0206.8--1309& 1.18& 0.320&   2.02 $\pm$\,  2.94& ...&   0.78 $\pm$\,
  1.24& ... \\ 
RXJ0258.7+0012 & 0.33& 0.259&   2.61 $\pm$\,  4.16& ...&   1.87 $\pm$\,
  1.69& ... \\ 
RXJ0259.5+0013 & 0.54& 0.194&   1.79 $\pm$\,  8.36& ...&   0.79 $\pm$\,
  0.84& ... \\ 
RXJ0351.6--3649\tablenotemark{\dagger}& 0.56& 0.372&   1.22 $\pm$\,  2.47& ...& ...& ... \\ 
RXJ0506.0--2840\tablenotemark{*}& 0.16& 0.136& ...& ...&   0.88 $\pm$\,
  1.01& ... \\ 
RXJ0521.1--2530\tablenotemark{\ddagger}& 2.57& 0.581&  -0.29 $\pm$\,  3.11&   0.60& ...& ... \\ 
RXJ0522.2--3625\tablenotemark{\dagger}& 1.79& 0.472&   4.83 $\pm$\,  5.85& ...& ...& ... \\ 
RXJ0826.4+3125 & 0.22& 0.209&  -4.98 $\pm$\,  9.83&   2.97&   0.90 $\pm$\,
  0.71& ... \\ 
RXJ0841.1+6422\tablenotemark{\dagger} & 1.49& 0.342&   4.91 $\pm$\, 10.11& ...& ...& ... \\ 
RXJ0842.8+5023\tablenotemark{\dagger} & 0.52& 0.423&   3.40 $\pm$\,  2.84& ...& ...& ... \\ 
RXJ0852.5+1618\tablenotemark{**} & 0.16& 0.098&  -2.33 $\pm$\,  3.12& ...&   1.17 $\pm$\,
  0.79& ... \\ 
RXJ0858.4+1357 & 0.70& 0.488&  -2.60 $\pm$\,  0.97&  11.28&   1.29 $\pm$\,
  1.14& ... \\ 
RXJ0907.2+3330\tablenotemark{\ddagger} & 0.48& 0.483&   1.58 $\pm$\,  4.76& ...& ...& ... \\ 
RXJ0921.2+4528 & 1.05& 0.315&  -1.74 $\pm$\,  6.75&   0.38&  -2.37 $\pm$\,
  4.65&   0.78 \\ 
RXJ0926.6+1242\tablenotemark{\dagger} & 1.75& 0.489&   3.81 $\pm$\,  3.64& ...& ...& ... \\ 
RXJ0943.7+1644\tablenotemark{**} & 0.31& 0.180&  -0.36 $\pm$\,  2.13& ...&   2.48 $\pm$\,
  1.11& ... \\ 
RXJ0958.2+5516\tablenotemark{*} & 0.97& 0.214& ...& ...&   1.47 $\pm$\,
  2.11& ... \\ 
RXJ1013.6+4933\tablenotemark{*} & 0.36& 0.133& ...& ...&   0.24 $\pm$\,
  0.56& ... \\ 
RXJ1015.1+4931\tablenotemark{\dagger} & 0.71& 0.383&   3.55 $\pm$\,  2.66& ...& ...& ... \\ 
RXJ1036.1+5713\tablenotemark{*} & 0.35& 0.203& ...& ...&   3.45 $\pm$\,
  1.89& ... \\ 
RXJ1049.0+5424 & 0.26& 0.251&   3.25 $\pm$\,  3.28& ...&   0.33 $\pm$\,
  0.78& ... \\ 
RXJ1117.2+1744 & 0.50& 0.305&   1.30 $\pm$\,  2.46& ...&  -3.16 $\pm$\,
  3.33&   1.24 \\ 
RXJ1117.5+1744\tablenotemark{\ddagger} & 1.90& 0.548&   0.48 $\pm$\,  1.14& ...& ...& ... \\ 
RXJ1120.9+2326\tablenotemark{\ddagger} & 2.89& 0.562&  -0.48 $\pm$\,  1.02&   0.98& ...& ... \\ 
RXJ1123.1+1409 & 0.93& 0.340&   2.72 $\pm$\,  3.54& ...&  -0.98 $\pm$\,
  3.70&   2.08 \\ 
RXJ1124.0--1700\tablenotemark{\dagger}& 0.81& 0.407&  -2.21 $\pm$\,  3.32&   4.58& ...& ... \\ 
RXJ1124.6+4155 \tablenotemark{*}& 0.67& 0.195& ...& ...&  -0.07 $\pm$\,
  0.87&   0.59 \\ 
RXJ1135.9+2131 & 0.14& 0.133&   2.26 $\pm$\,  5.48& ...&   0.23 $\pm$\,
  0.51& ... \\ 
RXJ1142.0+2144\tablenotemark{*} & 0.35& 0.131& ...& ...&   0.76 $\pm$\,
  0.82& ... \\ 
RXJ1146.4+2854 & 0.38& 0.149&   1.27 $\pm$\, 10.24& ...&   0.51 $\pm$\,
  0.60& ... \\ 
RXJ1158.1+5521 & 0.04& 0.135&  -2.74 $\pm$\, 14.32&   0.64&   0.47 $\pm$\,
  0.37& ... \\ 
RXJ1200.9--0327\tablenotemark{\dagger}& 1.28& 0.396&   1.05 $\pm$\,  3.23& ...& ...& ... \\ 
RXJ1206.5--0744\tablenotemark{*}& 0.26& 0.068& ...& ...&   0.20 $\pm$\,
  0.34& ... \\ 
RXJ1213.5+0253\tablenotemark{\dagger} & 1.07& 0.409&   4.21 $\pm$\,  8.94& ...& ...& ... \\ 
RXJ1218.4+3011 & 0.33& 0.368&  -4.16 $\pm$\,  3.98&   8.61&   8.25 $\pm$\,
  3.41& ... \\ 
RXJ1221.4+4918\tablenotemark{\ddagger} & 4.27& 0.700&   0.45 $\pm$\,  0.50& ...& ...& ... \\ 
RXJ1237.6+2632 & 0.25& 0.278&  -5.23 $\pm$\,  3.76&   7.38&  -0.53 $\pm$\,
  1.78&   1.95 \\ 
RXJ1254.6+2545\tablenotemark{**} & 0.17& 0.193&  -0.87 $\pm$\,  2.92& ...&   3.40 $\pm$\,
  1.02& ... \\ 
RXJ1254.8+2550\tablenotemark{*} & 0.32& 0.233& ...& ...&   0.71 $\pm$\,
  0.36& ... \\ 
RXJ1256.0+2556 & 0.24& 0.232&   1.08 $\pm$\,  4.54& ...&   0.16 $\pm$\,
  0.81& ... \\ 
RXJ1301.7+1059\tablenotemark{*} & 0.66& 0.231& ...& ...&   2.16 $\pm$\,
  1.27& ... \\ 
RXJ1309.9+3222\tablenotemark{**} & 0.34& 0.290&  -0.62 $\pm$\,  2.95& ...&  -0.55 $\pm$\,
  2.43& ... \\ 
RXJ1337.8+3854\tablenotemark{*} & 0.41& 0.252& ...& ...&   0.13 $\pm$\,
  1.22& ... \\ 
RXJ1342.8+4028 & 1.62& 0.699\tablenotemark{\ddagger}&   1.03 $\pm$\,  1.10& ...& ...& ... \\ 
RXJ1343.4+5547\tablenotemark{*} & 0.04& 0.069& ...& ...&   0.06 $\pm$\,
  0.25& ... \\ 
RXJ1354.8+6917 & 0.13& 0.207&   9.35 $\pm$\,  4.39& ...&  -2.02 $\pm$\,
  0.92&   3.42 \\ 
RXJ1438.9+6423\tablenotemark{*} & 0.25& 0.146& ...& ...&  -1.38 $\pm$\,
  0.49&   3.03 \\ 
RXJ1500.0+2233 & 0.35& 0.230&  -2.66 $\pm$\,  4.30&   3.69&   0.08 $\pm$\,
  0.74& ... \\ 
RXJ1515.5+4346\tablenotemark{*} & 0.29& 0.137& ...& ...&  -0.04 $\pm$\,
  0.50&   0.21 \\ 
RXJ1515.6+4350 & 0.28& 0.243&   7.76 $\pm$\,  7.80& ...&  -0.48 $\pm$\,
  0.71&  13.99 \\ 
RXJ1524.6+0957\tablenotemark{\ddagger} & 3.45& 0.516&  -0.72 $\pm$\,  1.36&   1.60& ...& ... \\ 
RXJ1537.7+1200 & 0.21& 0.134&  -4.22 $\pm$\, 10.94&   2.08&   0.62 $\pm$\,
  0.32& ... \\ 
RXJ1540.8+1445\tablenotemark{\ddagger} & 0.68& 0.441&   7.24 $\pm$\,  3.52& ...& ...& ... \\ 
RXJ1547.3+2056 & 0.79& 0.266&   1.84 $\pm$\,  4.29& ...&   3.01 $\pm$\,
  0.84& ... \\ 
RXJ1552.2+2013\tablenotemark{**} & 0.40& 0.136&   1.94 $\pm$\,  2.08& ...&   1.32 $\pm$\,
  0.70& ... \\ 
RXJ1630.2+2434\tablenotemark{*} & 0.33& 0.066& ...& ...&   1.19 $\pm$\,
  0.40& ... \\ 
RXJ1642.6+3935\tablenotemark{\dagger} & 0.58& 0.355&  -6.54 $\pm$\,  8.92&  22.03& ...& ... \\ 
RXJ1659.7+3410 & 0.52& 0.341&  -5.57 $\pm$\,  4.15&  10.64&   1.04 $\pm$\,
  4.64& ... \\ 
RXJ1722.8+4105 & 1.23& 0.309&   6.08 $\pm$\,  2.80& ...&   0.03 $\pm$\,
  3.10& ... \\ 
RXJ1729.0+7440 & 0.35& 0.213&  -1.02 $\pm$\, 11.77&   0.84&   1.94 $\pm$\,
  1.88& ... \\ 
RXJ1746.4+6848\tablenotemark{**} & 0.47& 0.217&  -6.83 $\pm$\,  4.53& ...&   0.53 $\pm$\,
  1.58& ... \\ 
RXJ2202.7--1902\tablenotemark{\ddagger}& 0.58& 0.438&  -6.93 $\pm$\,  8.21&  22.77& ...& ... \\ 
RXJ2212.6--1713& 0.04& 0.134&  -0.28 $\pm$\,  4.04&   0.24&   0.64 $\pm$\,
  0.22& ... \\ 
RXJ2213.5--1656& 0.71& 0.297& -13.55 $\pm$\,  9.81&  11.18&  -2.82 $\pm$\,
  2.86&  10.26 \\ 
RXJ2257.8+2056 & 0.44& 0.297&  -4.48 $\pm$\,  4.62&  13.26&   4.03 $\pm$\,
  1.72& ... \\ 
RXJ2328.8+1453\tablenotemark{\ddagger} & 0.85& 0.497&   4.47 $\pm$\,  2.07& ...& ...& ... \\ 
RXJ2348.8--3117& 0.49& 0.184&   8.86 $\pm$\, 10.67& ...&   1.16 $\pm$\,
  0.85& ... \\ 

\enddata

\tablenotetext{\Lambda}{For the luminosity calculations of \oii\, and
\ha\, the $\rm \Lambda CDM$ cosmology with $ \rm H_o=70\,km\,s^{-1}\,Mpc^{-1}$
was used.}
\tablenotetext{\dagger}{\ewha\, values excluded because of high noise at edge of spectrum}
\tablenotetext{\ddagger}{\ewha\, values excluded because line exceeds spectral coverage}
\tablenotetext{*}{\ewoii\, values excluded because of high noise at edge of spectrum}
\tablenotetext{**}{Emission-line luminosities excluded because spectra are not flux calibrated}
\tablenotetext{sky}{\ha\, line is coincident with an $\rm O_2$ sky absorption line}
\label{tab:tabem}
\end{deluxetable}

\begin{deluxetable}{cccc}
\tablecolumns{4}
\tablewidth{0pt}
\small
\tablecaption{The absorption line strengths for the BCGs.}
\tablehead{\colhead{Object} & \colhead{W$_{\rm H\beta}$} &
\colhead{W$_{\rm MgIb}$} & \colhead{I$_{\rm Mg_2}$} \\  \colhead{} &
\colhead{(\AA)} & \colhead{(\AA)} & \colhead{(mag)}}
\startdata

RXJ0041.1--2339&   2.24 $\pm$\,  0.64&   3.30 $\pm$\,  0.94&   0.26 $\pm$\,
  0.11 \\
RXJ0050.9--0929&   1.07 $\pm$\,  0.42&   2.97 $\pm$\,  0.73&   0.31 $\pm$\,
  0.07 \\
RXJ0056.9--2213&   2.32 $\pm$\,  0.90&   3.79 $\pm$\,  0.72&   0.29 $\pm$\,
  0.13 \\
RXJ0110.3+1938\tablenotemark{\ddagger} &   1.90 $\pm$\,  0.85&   ...&   0.27 $\pm$\,
  0.10 \\
RXJ0122.5--2832&   2.32 $\pm$\,  0.41&   3.95 $\pm$\,  0.60&   0.29 $\pm$\,
  0.11 \\
RXJ0124.5+0400 &   2.17 $\pm$\,  0.92&   4.19 $\pm$\,  1.03&   0.27 $\pm$\,
  0.14 \\
RXJ0142.8+2025 &   2.24 $\pm$\,  0.46&   3.09 $\pm$\,  0.91&   0.24 $\pm$\,
  0.12 \\
RXJ0144.4+0212 &   0.61 $\pm$\,  0.47&  -0.54 $\pm$\,  0.78&   0.21 $\pm$\,
  0.06 \\
RXJ0159.3+0030 &   1.63 $\pm$\,  1.00&   0.05 $\pm$\,  0.69&   0.14 $\pm$\,
  0.08 \\
RXJ0206.3+1511 &   1.11 $\pm$\,  1.02&   3.86 $\pm$\,  1.21&   0.25 $\pm$\,
  0.16 \\
RXJ0206.8--1309\tablenotemark{\ddagger}&   0.51 $\pm$\,  1.43&   ...&   0.28 $\pm$\,
  0.11 \\
RXJ0258.7+0012 &   2.32 $\pm$\,  1.33&   1.93 $\pm$\,  1.30&   0.25 $\pm$\,
  0.16 \\
RXJ0259.5+0013 &   2.38 $\pm$\,  0.76&   5.00 $\pm$\,  1.11&   0.25 $\pm$\,
  0.14 \\
RXJ0351.6--3649&   2.29 $\pm$\,  0.90&   4.78 $\pm$\,  0.99&   0.22 $\pm$\,
  0.18 \\
RXJ0506.0--2840&   2.37 $\pm$\,  1.13&   0.13 $\pm$\,  1.79&   0.22 $\pm$\,
  0.12 \\
RXJ0521.1--2530&   1.13 $\pm$\,  1.77&   5.99 $\pm$\,  3.26&   0.38 $\pm$\,
  0.41 \\
RXJ0522.2--3625\tablenotemark{\ddagger}&   1.63 $\pm$\,  1.34& ...& 0.60 $\pm$\, 0.60 \\
RXJ0826.4+3125 &   2.54 $\pm$\,  0.99&   3.25 $\pm$\,  0.75&   0.24 $\pm$\,
  0.11 \\
RXJ0841.1+6422 &  -0.99 $\pm$\,  1.64&   5.74 $\pm$\,  2.51&   0.25 $\pm$\,
  0.28 \\
RXJ0842.8+5023 &   1.98 $\pm$\,  1.75&   3.20 $\pm$\,  2.04&   0.20 $\pm$\,
  0.29 \\
RXJ0852.5+1618 &   1.57 $\pm$\,  1.38&   4.56 $\pm$\,  1.41&   0.34 $\pm$\,
  0.24 \\
RXJ0858.4+1357 &   1.68 $\pm$\,  0.45&   2.55 $\pm$\,  0.79&   0.19 $\pm$\,
  0.11 \\
RXJ0907.2+3330 &   2.41 $\pm$\,  0.92&   1.90 $\pm$\,  1.56&   0.24 $\pm$\,
  0.25 \\
RXJ0921.2+4528 &   0.82 $\pm$\,  1.22&   1.35 $\pm$\,  0.84&   0.16 $\pm$\,
  0.19 \\
RXJ0926.6+1242\tablenotemark{*} &   3.71 $\pm$\,  2.61& ...&   0.18 $\pm$\,
  0.51 \\
RXJ0943.7+1644 &   2.59 $\pm$\,  0.76&   4.97 $\pm$\,  1.58&   0.31 $\pm$\,
  0.25 \\
RXJ0958.2+5516 &  -0.33 $\pm$\,  1.84&   7.49 $\pm$\,  1.70&   0.34 $\pm$\,
  0.28 \\
RXJ1013.6+4933 &   1.49 $\pm$\,  1.38&   3.99 $\pm$\,  1.03&   0.32 $\pm$\,
  0.13 \\
RXJ1015.1+4931 &   1.77 $\pm$\,  1.48&  -0.15 $\pm$\,  0.95&   0.22 $\pm$\,
  0.17 \\
RXJ1036.1+5713 &   6.68 $\pm$\,  1.44&   1.55 $\pm$\,  1.31&   0.08 $\pm$\,
  0.13 \\
RXJ1049.0+5424 &   2.64 $\pm$\,  0.36&   1.59 $\pm$\,  0.67&   0.16 $\pm$\,
  0.10 \\
RXJ1117.2+1744 &   2.24 $\pm$\,  1.43&   1.42 $\pm$\,  1.66&   0.03 $\pm$\,
  0.12 \\
RXJ1117.5+1744\tablenotemark{\dagger} & ...&   2.88 $\pm$\,  0.68&   0.24 $\pm$\,
  0.16 \\
RXJ1120.9+2326\tablenotemark{\dagger} & ...&   1.17 $\pm$\,  0.77&   0.28 $\pm$\,
  0.14 \\
RXJ1123.1+1409 &   2.09 $\pm$\,  0.93&   1.38 $\pm$\,  1.58&   0.23 $\pm$\,
  0.14 \\
RXJ1124.0--1700&  -0.32 $\pm$\,  1.65&   0.43 $\pm$\,  1.24&   0.17 $\pm$\,
  0.23 \\
RXJ1124.6+4155 &   3.88 $\pm$\,  1.19&   2.72 $\pm$\,  0.92&   0.20 $\pm$\,
  0.11 \\
RXJ1135.9+2131 &   0.34 $\pm$\,  0.54&   2.67 $\pm$\,  0.48&   0.24 $\pm$\,
  0.11 \\
RXJ1142.0+2144 &   4.21 $\pm$\,  1.50&   3.61 $\pm$\,  1.31&   0.41 $\pm$\,
  0.16 \\
RXJ1146.4+2854 &   0.80 $\pm$\,  0.79&   1.25 $\pm$\,  0.42&   0.23 $\pm$\,
  0.14 \\
RXJ1158.1+5521 &   1.56 $\pm$\,  0.80&   2.48 $\pm$\,  0.65&   0.26 $\pm$\,
  0.10 \\
RXJ1200.9--0327&   0.58 $\pm$\,  1.33&   2.53 $\pm$\,  1.71&   0.22 $\pm$\,
  0.17 \\
RXJ1206.5--0744\tablenotemark{*}& ...& ...&   0.05 $\pm$\,
  0.05 \\
RXJ1213.5+0253 &   4.27 $\pm$\,  1.39&   3.05 $\pm$\,  1.87&   0.19 $\pm$\,
  0.16 \\
RXJ1218.4+3011 &   2.05 $\pm$\,  1.08&   4.73 $\pm$\,  1.60&   0.22 $\pm$\,0.21 \\
RXJ1221.4+4918 &  -0.18 $\pm$\,  0.44&   6.87 $\pm$\,  0.99&   0.13 $\pm$\,0.28 \\
RXJ1237.6+2632 &  -2.36 $\pm$\,  0.82&   3.81 $\pm$\,  1.22&   0.16 $\pm$\,
  0.11 \\
RXJ1254.6+2545 &  -0.28 $\pm$\,  1.13&   4.47 $\pm$\,  1.33&   0.23 $\pm$\,
  0.25 \\
RXJ1254.8+2550 &   1.69 $\pm$\,  0.89&   3.26 $\pm$\,  0.73&   0.24 $\pm$\,
  0.08 \\
RXJ1256.0+2556 &   1.00 $\pm$\,  0.69&   3.40 $\pm$\,  0.84&   0.24 $\pm$\,
  0.13 \\
RXJ1301.7+1059 &   1.88 $\pm$\,  2.13&   4.50 $\pm$\,  1.89&   0.30 $\pm$\,
  0.16 \\
RXJ1309.9+3222 &   2.71 $\pm$\,  1.45&   4.64 $\pm$\,  2.24&   0.21 $\pm$\,
  0.36 \\
RXJ1337.8+3854 &  -1.68 $\pm$\,  1.36&   0.60 $\pm$\,  1.08&   0.24 $\pm$\,
  0.11 \\
RXJ1342.8+4028 &   0.50 $\pm$\,  0.95&   3.69 $\pm$\,  1.72&   0.28 $\pm$\, 0.36\\
RXJ1343.4+5547 &   0.90 $\pm$\,  0.38&   3.29 $\pm$\,  0.81&   0.27 $\pm$\,
  0.10 \\
RXJ1354.8+6917 &  -1.15 $\pm$\,  0.75&   3.73 $\pm$\,  1.33&   0.40 $\pm$\,
  0.17 \\
RXJ1438.9+6423 &   3.32 $\pm$\,  0.69&   2.30 $\pm$\,  0.90&   0.25 $\pm$\,
  0.15 \\
RXJ1500.0+2233 &   1.93 $\pm$\,  0.97&   1.22 $\pm$\,  1.20&   0.15 $\pm$\,
  0.11 \\
RXJ1515.5+4346\tablenotemark{\dagger} &   2.42 $\pm$\,  1.85&   3.63 $\pm$\,  1.43&   0.25 $\pm$\,
  0.10 \\
RXJ1515.6+4350 &   1.08 $\pm$\,  0.58&   2.96 $\pm$\,  0.74&   0.21 $\pm$\,
  0.10 \\
RXJ1524.6+0957 &   2.18 $\pm$\,  0.64&   3.09 $\pm$\,  1.17&   0.17 $\pm$\,
  0.17 \\
RXJ1537.7+1200 &   1.84 $\pm$\,  0.49&   2.07 $\pm$\,  0.72&   0.25 $\pm$\,
  0.06 \\
RXJ1540.8+1445 &   1.61 $\pm$\,  1.96&   6.07 $\pm$\,  2.14&   0.10 $\pm$\, 0.38 \\
RXJ1547.3+2056 &   0.79 $\pm$\,  0.72&   1.01 $\pm$\,  1.15&   0.26 $\pm$\,
  0.12 \\
RXJ1552.2+2013 &   1.88 $\pm$\,  1.18&   5.70 $\pm$\,  1.22&   0.32 $\pm$\,
  0.20 \\
RXJ1630.2+2434 &   1.41 $\pm$\,  0.86&   4.42 $\pm$\,  0.49&   0.38 $\pm$\,
  0.17 \\
RXJ1642.6+3935 &  -0.41 $\pm$\,  2.74&  -0.90 $\pm$\,  1.71&   0.03 $\pm$\,
  0.29 \\
RXJ1659.7+3410 &   1.34 $\pm$\,  0.86&   4.24 $\pm$\,  0.88&   0.27 $\pm$\,
  0.16 \\
RXJ1722.8+4105 &  -0.26 $\pm$\,  1.36&   2.47 $\pm$\,  1.90&   0.32 $\pm$\, 0.22 \\
RXJ1729.0+7440 &   5.10 $\pm$\,  2.00&  -0.44 $\pm$\,  2.57&   0.23 $\pm$\, 0.36 \\
RXJ1746.4+6848 &   0.04 $\pm$\,  2.03&   2.22 $\pm$\,  1.84&   0.23 $\pm$\,
  0.25 \\
RXJ2202.7--1902&   2.19 $\pm$\,  1.11&   3.43 $\pm$\,  1.24&   0.22 $\pm$\,
  0.20 \\
RXJ2212.6--1713&   1.45 $\pm$\,  0.57&   3.06 $\pm$\,  0.44&   0.27 $\pm$\,
  0.09 \\
RXJ2213.5--1656&   2.15 $\pm$\,  1.35&   4.26 $\pm$\,  1.06&   0.26 $\pm$\,
  0.15 \\
RXJ2257.8+2056 &   2.62 $\pm$\,  0.66&   4.04 $\pm$\,  1.04&   0.24 $\pm$\,
  0.10 \\
RXJ2328.8+1453 &  -1.44 $\pm$\,  1.15&   3.94 $\pm$\,  2.30&   0.30 $\pm$\, 0.24 \\
RXJ2348.8--3117&   1.05 $\pm$\,  0.38&   3.15 $\pm$\,  0.92&   0.23 $\pm$\,
  0.14 \\

\enddata
\label{tab:tabab}
\tablenotetext{\dagger}{\hb\, line is coincident with an $\rm O_2$ sky absorption line}
\tablenotetext{\ddagger}{\mgib\, is coincident with an $\rm O_2$ sky absorption line}
\tablenotetext{*}{\mgib\, spectrum show local disturbance}
\end{deluxetable}

\begin{deluxetable}{ccccc}
\tablewidth{0pt}
\tablecolumns{5}
\tablecaption{The number of BCGs per X-ray luminosity bin for the BCS and this sample, as well as the frequency of detection for BCGs in the BCS.}
\tablehead{\colhead{$\rm L_X$ Range} & \colhead{160SD} & \colhead{BCS} & \colhead{BCS} & \colhead{Fraction Detected} \\
\colhead{(\lum)} & \colhead{BCGs}  & \colhead{BCGs} &
 \colhead{Emission-line} & \colhead{in BCS}\\
\colhead{} & \colhead{} & \colhead{} & \colhead{BCGs\tablenotemark{\dagger}} 
& \colhead{}}
\startdata
  $\rm L_X \geq 10^{45}$                    &0     &33   &10  &30\%\\
  $\rm 10^{44} \leq L_X \leq 10^{45}$       &29    &103  &18  &18\%\\
  $\rm 10^{42} \leq L_X \leq 10^{44}$       &48    &31   &5   &16\%\\
\enddata
\label{tab:emcomp}

\tablenotetext{\dagger}{Emission-line BCGs with equivalent width
estimates greater than $\sim-20$\AA\, for \oii\, and $\sim-10$\AA\,
for \ha.}

\end{deluxetable}

\end{document}